\documentclass[twocolumn,english,twocolumn,prb,showpacs]{revtex4}
\usepackage[latin9]{inputenc}
\usepackage{graphicx,color}
\usepackage{amssymb}
\usepackage[bookmarks]{hyperref}

\newcommand{\beq}{\begin{equation}}
\newcommand{\eeq}{\end{equation}}

\begin{document}

\title{The Crooks relation and its connection to the detailed balance principle of equilibrium correlation functions}

\author{M. Heyl}

\author{S. Kehrein}

\affiliation{Department of Physics, Arnold Sommerfeld Center for Theoretical Physics,
and Center for NanoScience, Ludwig-Maximilians-Universit\"at M\"unchen,
Theresienstr. 37, 80333 Munich, Germany}

\begin{abstract}
	We show that in the quantum case any work distribution can be related to an equilibrium correlation function in an extended Hilbert space. As a consequence of this identification the Crooks relation is a restatement of the detailed balance principle for equilibrium correlation functions. The presented derivation serves as an alternative proof of the Crooks relation residing only on the detailed balance principle.
\end{abstract}


\maketitle

\emph{Introduction.-}
Within the field of nonequilibrium physics the non-equilibrium work fluctuation theorems such as the Crooks relation~\cite{Crooks} and the Jarzynski equality\cite{Jarzynski} establish one of the rare universal relations in systems that are driven out of equilibrium. Remarkably, these theorems are independent of the microscopic details apart from the knowledge of equilibrium free energies. Moreover, they are independent of the nonequilibrium process even if the system is driven far away from equilibrium such that the range of their validity extends way beyond the linear response regime.

If a system is driven out of equilibrium via some arbitrary process work $\omega$ is performed on this system with a probability distribution function $P_F(\omega)$ which is also called work distribution~\cite{Talkner_work}. The Crooks relation establishes a general connection between $P_F(\omega)$ and the work distribution $P_B(\omega)$ for the time-reversed protocol~\cite{Crooks,Tasaki, Talkner_Crooks, Talkner_Crooks_open}
\beq
	\frac{P_F(\omega)}{P_B(-\omega)}=e^{\beta(\omega-\Delta F)}
	\label{eq_Crooks}
\eeq
if for the forward and backward process the systems are initially prepared in thermal states with the same inverse temperature $\beta$. The system details enter only via the equilibrium free energy difference $\Delta F$ between the free energies of the two initial equilibrium ensembles of the forward and backward process. Remarkably, the Crooks relation is independent of the actual nonequilibrium protocol and only depends on equilibrium quantities. The Crooks relation has been proven for classical systems~\cite{Crooks} as well as for quantum systems~\cite{Tasaki,Talkner_Crooks,Talkner_Crooks_open}. The Jarzynksi equality~\cite{Jarzynski} is the integrated version of Eq.~(\ref{eq_Crooks}), see, e.g., Ref.~\cite{Crooks}. In the case where the system is prepared in a microcanonical state the Crooks relation is still valid, however, entropy differences appear instead of free energy differences~\cite{Talkner_micro}.

For classical systems the Crooks relation has been observed experimentally~\cite{RNA,Crooks_exp} and used to measure equilibrium free energy differences between the folded and unfolded state of RNA-hairpins~\cite{RNA}. In the quantum case, however, the Crooks relation has yet not been observed experimentally due to the difficulty of measuring its fundamental ingredient, the work distribution. However, there exist proposals within the scope of current experimental technology that potentially allow for an experimental investigation~\cite{Heyl,Huber}.

In this work we show that the universality of the Crooks relation for quantum systems can be traced back to a universal relation in thermal equilibrium states, namely the detailed balance principle for correlation functions. For classical systems the Crooks relation has been proven on the basis  of a detailed balance relation between the probabilities of time-reversed paths in phase space, see Ref.~\cite{Crooks}. After introducing the basic notions we will prove the connection between the Crooks relation and detailed balance for quantum systems in the remainder of this letter.

\begin{figure}
\centering
\includegraphics[width=0.8\columnwidth]{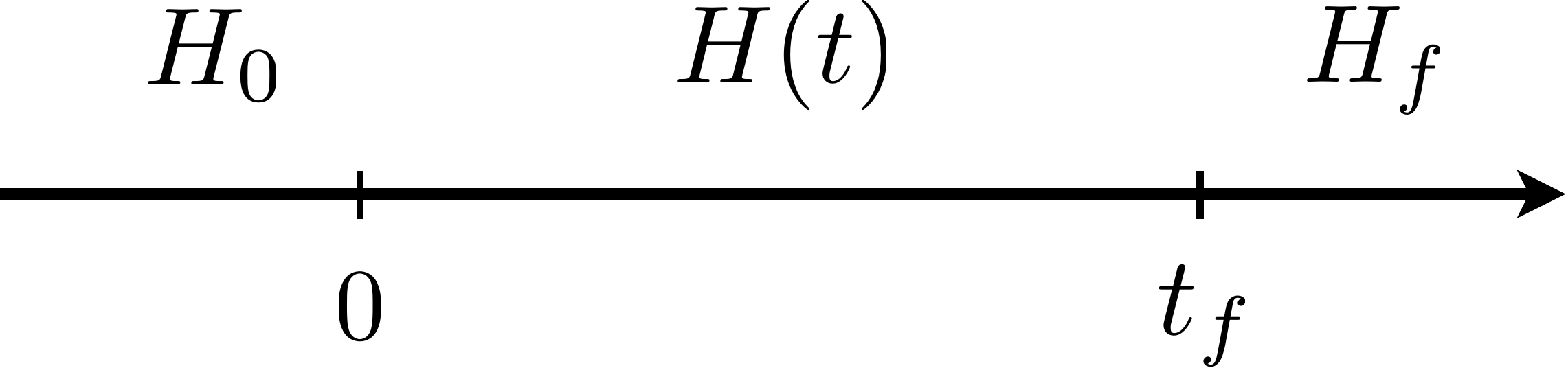}
\caption{Schematic picture of the time-dependent setup. For times $t<0$ the system is prepared in a canonical state at inverse temperature $\beta$ with a Hamiltonian $H(t<0)=H_0$. At time $t=0$ some arbitrary protocol starts and the system Hamiltonian $H(t)$ becomes time-dependent up to a time $t=t_f$ where the protocol ends and the system Hamiltonian is given by $H(t>t_f)=H_f$.}
\label{Fig1}
\end{figure}

\emph{Reduction to a quench problem.-}
Consider a closed quantum system described by a Hamiltonian $H_0$ prepared in a thermal state at inverse temperature $\beta$. As we require that the system is prepared in a canonical state we implicitely assume that the system is connected to a thermostat. If the coupling, however, is weak the influence of the heat bath apart from establishing a thermal state for the system can be neglected~\cite{Talkner_open_weak}. Note that the Crooks relation is valid also for open quantum systems even if the coupling to the environment is strong~\cite{Talkner_Crooks_open}. At time $t=0$ we start a protocol under which the Hamiltonian $H=H(t)$ acquires a time-dependence up to a finite time $t=t_f$ where the protocol stops, see Fig.~\ref{Fig1} for an illustration. For times $t>t_f$ the system Hamiltonian is denoted by $H_f$. Due to the time-dependence of the Hamiltonian energy is no longer a constant of motion such that work is performed on the system. It is, however, impossible to define a corresponding work operator as the work performed requires two energy measurements~\cite{Tasaki,Talkner_work}. This is true apart from the exceptional case $\beta\to\infty$ where the energy of the initial state, the ground state of $H(t<0)$, is known precisely. Due to the inherent randomness in the initial thermal state at finite $\beta$ the work $\omega$ rather is a random variable with a probability distribution function~\cite{Talkner_work}
\beq
	P_F(\omega)=\int \frac{ds}{2\pi} \: e^{i \omega s} G_F(s)
\eeq
that  via Fourier transformation is related to a dynamical correlation function $G_F(s)$~\cite{Talkner_work}
\begin{eqnarray}
	G_F(s)=\frac{1}{Z_F} \mathrm{Tr} \left( e^{-\beta H_0} e^{i H_0 s} U_F^\dag (t_f) e^{-i H_f s} U_F(t_f) \right). 
\end{eqnarray}
Here $Z_F=\mathrm{Tr}\: e^{-\beta H_0}$ is the partition function corresponding to the Hamiltonian $H_0$ at inverse temperature $\beta$. The details of the protocol enter $P_F(\omega)$ via the time evolution operator $U_F(t_f)=\mathrm{T} \exp\left[ -i \int_0^{t_f} dt \: H(t) \right]$ from time $t=0$ up to time $t=t_f$ with $\mathrm{T}$ the usual time ordering prescription. Here and in the rest of the letter we set $\hbar=1$. The work distribution for the backward process is obtained analogously
\begin{eqnarray}
	& & P_B(\omega)=\int\frac{ds}{2\pi} \: e^{i\omega s} \: G_B(s) \nonumber \\
	G_B(s) &=&\frac{1}{Z_B} \mathrm{Tr} \left( e^{-\beta H_f} e^{i H_f s} U_F(t_f) e^{-H_0 s} U_F^\dag(t_f) \right)
\end{eqnarray}
due to the relation $U_B(t_f)=U_F^\dag(t_f)$ for time-reversed processes. The quantity $Z_B=\mathrm{Tr} e^{-\beta H_f}$ denotes the partition function for the Hamiltonian $H_f$.

If the protocol only consists of an aprupt change at time $t=0$, i.e., $H(t<0)=H_0$ and $H(t>0)=H_f$, a socalled quench, we have $U_B(t_f)=U_F(t_f)=1$. The more general case shown above can be reduced to such a quench problem if we define $H_0:=H(0)$ and $\tilde{H}_f:=U_F^\dag(t_f) H(t_f) U_F(t_f)$:
\begin{eqnarray}
	G_F(s)=\frac{1}{Z_F} \mathrm{Tr} \left( e^{-\beta H_0} e^{i H_0 s} e^{-i \tilde{H}_f s}  \right), \nonumber \\
	G_B(s)=\frac{1}{Z_B} \mathrm{Tr} \left( e^{-\beta \tilde{H}_f} e^{i \tilde{H}_f s} e^{-i H_0 s}  \right).
\end{eqnarray}
Thus, we can concentrate on the problem of a sudden perturbation without restriction, the Hamiltonian $\tilde{H}_f$, however, will be a very complicated object for a nontrivial protocol. This representation is therefore not suitable for concrete applications, for the following analysis it proves to be very useful. Note that $Z_B=\mathrm{Tr} \: e^{-\beta H_f}=\mathrm{Tr}\: e^{-\beta \tilde{H}_f}$. For later convenience we define $V:=\tilde{H}_f-H_0$ such that
\beq
	\tilde{H}_f=H_0+V.
\eeq
The Hamiltonian $\tilde{H}_f$ can thus be interpreted to consist of a free part $H_0$ and a perturbation $V$ that, however, can be arbitrarily strong and very complicated depending on the details of the protocol.

\emph{Detailed balance.-}
To prove the connection between the Crooks relation and the detailed balance principle it is suitable to introduce an artificial fermionic degree of freedom with a corresponding creation, $b^\dag$, and annihilation operator, $b$. In the extended Hilbert space we define a new Hamiltonian
\beq
	\mathcal{H}=H_0+V bb^\dag
\eeq
containing both the initial degrees of freedom as well as the new artificial fermion. Note that the single-particle energy $\varepsilon_b=0$ of the $b$-fermion vanishes if we neglect the perturbation $V bb^\dag$ such that the creation or annihilation of the $b$-fermion costs zero energy. Moreover, the Hamiltonian $\mathcal{H}$ conserves the number of $b$-fermions which will be important in what follows. The Hamiltonian $\mathcal{H}$ has a different structure in the occupied and unoccupied sectors for the $b$-fermion. If the orbital $b$ is occupied we have $\mathcal{H}\to H_0$, if it is unoccupied we have $\mathcal{H} \to H_0+V$. The equilibrium two-point correlation functions for the artificial fermion with respect to the Hamiltonian $\mathcal{H}$ equal
\begin{eqnarray}
	\langle b^\dag(s) b \rangle = \frac{1}{Z} \mathrm{Tr} \left( e^{-\beta \mathcal{H}} b^\dag(s) b \right) \nonumber\\
	\langle b(s) b^\dag \rangle = \frac{1}{Z} \mathrm{Tr} \left( e^{-\beta \mathcal{H}} b(s) b^\dag \right) 
\end{eqnarray}
with $Z=\mathrm{Tr} e^{-\beta \mathcal{H}}=Z_F+Z_B$ and $b(s)=e^{i \mathcal{H} s} b e^{-i \mathcal{H} s}$. Using the elementary property $bb=b^\dag b^\dag=0$ for fermionic operators it is straightforward to show that
\beq
	\langle b^\dag(s) b \rangle= \frac{Z_F}{Z} G_F(s),\: \langle b(s) b^\dag \rangle= \frac{Z_B}{Z} G_B(s).
\label{eq_b_G}
\eeq
The artificial $b$-fermion is used as a formal device to implement the switch on and off of the ``perturbation`` $V$. Moreover, the equation above shows that the generating function of any work distribution is proportional to an equilibrium correlation function. Introducing the Fourier transforms, e.g., $\langle b^\dag b \rangle_\omega=(2\pi)^{-1}\int ds \: e^{i \omega s} \langle b^\dag(s) b\rangle$, of the $b$-fermion correlation functions the detailed balance principle relates $\langle b^\dag b\rangle_\omega$ and $\langle b b^\dag\rangle_\omega$ via
\beq
	\frac{\langle b^\dag b \rangle_\omega}{\langle b b^\dag \rangle_{-\omega}}=e^{\beta \omega}.
\eeq
Plugging in Eq.~(\ref{eq_b_G}) one directly proves the Crooks relation. Thus, we have shown that the Crooks relation can be related to the detailed balance principle via an artificial fermionic degree of freedom. Moreover, the derivation presented above can be considered as an alternative proof of the Crooks relation that only resorts to an elementary equilibrium property, the detailed balance principle. Note that this construction follows the spirit of the X-ray edge problem where, however, the $b$-particle has a physical content as a deep lying electronic core state in a metal~\cite{X_ray_edge}. In the X-ray edge problem the $b$-fermion switches on and off a potential scatterer for conduction band electrons. The relation between the X-ray edge problem and work distributions and thus the nonequilibrium work fluctuation theorems has been worked out recently~\cite{Heyl}.

\emph{Conclusion.-}
In this work we have shown that work distributions can be identified with equilibrium correlation functions in an extended Hilbert space with an additional artificial fermionic degree of freedom. Within this identification the Crooks relation is a restatement of the detailed balance principle of equilibrium correlation functions. As a consequence we have shown an alternative proof of the Crooks relation for quantum systems that is based on the detailed balance property of correlation functions in thermal states.

\emph{Acknowledgements.-} We acknowledge stimulating discussions with Constantin Tomaras. This work was supported by SFB~TR12 of the Deutsche Forschungsgemeinschaft (DFG), the Center for Nanoscience (CeNS) Munich, and the German Excellence Initiative via the Nanosystems Initiative Munich (NIM).

\bibliographystyle{apsrev}

\end{document}